\documentclass[manuscript]{emulateapj}
\usepackage{apjfonts}
\usepackage{ulem}
\usepackage[usenames,dvipsnames]{color}

\newcommand{\mldyn}{\ifmmode\Upsilon_\mathrm{dyn}\else$\Upsilon_\mathrm{dyn}$\fi}
\newcommand{\mldynK}{\ifmmode\Upsilon_\mathrm{dyn}^K\else$\Upsilon_\mathrm{dyn}^K$\fi}
\newcommand{\mldynn}{\ifmmode\Upsilon_{\mathrm{dyn},0}\else$\Upsilon_{\mathrm{dyn},0}$\fi}
\newcommand{\mlstar}{\ifmmode\Upsilon_\ast\else$\Upsilon_\ast$\fi}
\newcommand{\mlstarK}{\ifmmode\Upsilon_\ast^K\else$\Upsilon_\ast^K$\fi}
\newcommand{\musigrel}{$\mu$-$\Sigma$-relation}
\newcommand{\hz}{\ifmmode h_z\else$h_z$\fi}
\newcommand{\hR}{\ifmmode h_R\else$h_R$\fi}
\newcommand{\hsig}{\ifmmode h_{\sigma,z}\else$h_{\sigma,z}$\fi}
\newcommand{\mlunit}{$M_\odot/L_\odot$}
\newcommand{\mun}{\ifmmode \mu_{K,0}\else$\mu_{K,0}$\fi}
\newcommand{\sd}{\ifmmode \Sigma\else$\Sigma$\fi}
\newcommand{\sdyn}{\ifmmode \Sigma_\mathrm{dyn}\else$\Sigma_\mathrm{dyn}$\fi}
\newcommand{\sdynn}{\ifmmode \Sigma_{\mathrm{dyn},0}\else$\Sigma_{\mathrm{dyn},0}$\fi}
\newcommand{\si}{\ifmmode \sigma_\mathrm{i}\else$\sigma_\mathrm{i}$\fi}
\newcommand{\slos}{\ifmmode\sigma_\mathrm{LOS}\else$\sigma_\mathrm{LOS}$\fi}
\newcommand{\sz}{\ifmmode\sigma_z\else$\sigma_z$\fi}
\newcommand{\sR}{\ifmmode\sigma_R\else$\sigma_R$\fi}
\newcommand{\szn}{\ifmmode\sigma_{z,0}\else$\sigma_{z,0}$\fi}
\newcommand{\etal}{et al.}

\slugcomment{Accepted for publication in ApJ Letters}

\shorttitle{The Link between Light and Mass in Galaxy Disks}
\shortauthors{Swaters et al.}

\begin{document}


\title{The Link Between Light and Mass in Late-type Spiral Galaxy Disks}

\author{
Robert A. Swaters,\altaffilmark{1}
Matthew A. Bershady,\altaffilmark{2}
Thomas P. K. Martinsson,\altaffilmark{3}
Kyle B. Westfall,\altaffilmark{4}
David R. Andersen,\altaffilmark{5} and
Marc A. W. Verheijen\altaffilmark{6}
}

\altaffiltext{1}{National Optical Astronomical Observatory,
950 North Cherry Ave, Tucson, AZ 85719, USA}

\altaffiltext{2}{University of Wisconsin, Department of Astronomy, 475
N. Charter St., Madison, WI 53706}

\altaffiltext{3}{Leiden Observatory, Leiden University, PO Box 9513,
  2300 RA Leiden, the Netherlands}

\altaffiltext{4}{Institute of Cosmology and Gravitation, University of
  Portsmouth, Dennis Sciama Building, Burnaby Road, Portsmouth PO1
  3FX, UK }

\altaffiltext{5}{NRC Herzberg Programs in Astronomy and Astrophysics, 5071 W
Saanich Road, Victoria, BC V9E 2E7, Canada}

\altaffiltext{6}{University of Groningen, Kapteyn Astronomical
  Institute, Landleven 12, 9747 AD Groningen, Netherlands}
\email{rob@swaters.net}

\begin{abstract}
We present the correlation between the extrapolated central disk
surface brightness ($\mu$) and extrapolated central surface mass
density ($\Sigma$) for galaxies in the DiskMass sample. This
\musigrel\ has a small scatter of 30\% at the high-surface-brightness
(HSB) end. At the low surface brightness (LSB) end, galaxies fall
above the \musigrel, which we attribute to their higher dark matter
content. After correcting for the dark matter, as well as for the
contribution of gas and the effects of radial gradients in the disk,
the LSB end falls back on the linear \musigrel. The resulting scatter
about the corrected \musigrel\ is 25\% at the HSB end, and about 50\%
at the LSB end. The intrinsic scatter in the \musigrel\ is estimated
to be 10\% to 20\%.  Thus, if \mun\ is known, the stellar surface mass
density is known to within 10-20\% (random error).  Assuming disks
have an exponential vertical distribution of mass, the average
\mlstarK\ is $0.24$~\mlunit, with an intrinsic scatter around the mean
of at most $0.05$~\mlunit.  This value for \mlstarK\ is 20\% smaller
than we found in Martinsson \etal, mainly due to the correction for
dark matter applied here. This small scatter means that among the
galaxies in our sample variations in scale height, vertical density
profile shape, and/or the ratio of vertical over radial velocity
dispersion must be small.
\end{abstract}


\section{Introduction}

Mass-modeling of rotation curves provided the first qualitative
indication that low surface-brightness disks were sub-maximal and had
lower densities (e.g., de Blok \& McGaugh 1997). This method, however,
is limited by the disk-halo degeneracy (van Albada et al. 1985),
whereby contributions of disk and halo can range from halo-only to a
maximum disk. Measurements of the vertical stellar dispersion of disk
galaxies provide a powerful tool to measure the disk surface mass
densities (Bahcall 1984), breaking this degeneracy.  Results based on
vertical stellar velocity dispersion measurements show that even
normal surface-brightness disks are significantly submaximal (Bottema
1993; Kregel \etal\ 2005; Bershady \etal\ 2011; Martinsson
\etal\ 2013b, hereafter DMS-VII), similar to values found from other
work, such as PNe kinematics (Herrmann \& Ciardullo 2009) and
gravitational lensing (Dutton \etal\ 2011); work based on
hydrodynamical modeling find higher values (Weiner \etal\ 2001; Kranz
\etal\ 2003).  Results for the Milky Way range from submaximal to
maximal, depending on the value of the derived radial scale length
(e.g., Sackett 1997; Bovy \& Rix (2013).

For a self-gravitating disk in
equilibrium the dynamical local surface mass density \sdyn\ can be
determined from:
\begin{equation}
\sdyn = I \mldyn = \sz^2/(\pi G k \hz),
\label{eqsigmaz}
\end{equation}
where \sz\ is the stellar vertical velocity dispersion, $k$ a constant
depending on the vertical mass distribution (1.5 for an exponential
distribution, 2 for an isothermal), \hz\ the vertical scale height,
$I$ the surface luminosity density, and \mldyn\ the dynamical
mass-to-light ratio of the disk. Thus, to determine the surface mass
density (\sd) in a galactic disk, both the vertical distribution of
stars and the vertical stellar velocity dispersion are needed (e.g.,
Bahcall 1984). To measure the vertical stellar velocity dispersion
without the uncertainties introduced by projection effects, face-on
galaxies are needed, and to measure the vertical distribution of stars
unambiguously, edge-on galaxies are needed. It is therefore not
possible to measure both simultaneously in external galaxies.

Fortunately, the relation between the scale height and scale length is
statistically well known from edge-on galaxies (see Bershady \etal\ 2010b;
hereafter DMS-II). Combining this knowledge about scale heights from
edge-on galaxies with the measured vertical stellar velocity
dispersion from nearly face-on galaxies, one can calculate \sdyn.

The DiskMass Survey (Bershady \etal\ 2010a, hereafter DMS-I) has been
designed to reliably measure surface brightness, inclination, and
vertical velocity dispersion simultaneously (DMS-II; Martinsson
\etal\ 2013a, hereafter DMS-VI). With these, Martinsson \etal\ (2013b)
derived dynamical and stellar surface mass densities, as well as the
stellar mass-to-light ratio \mlstarK. Mass modeling based on these
\mlstarK\ show disks to be significantly submaximal, with the disks
contributing on average $57\%\pm 7$\% of the rotation velocity at 2.2
disk scale lengths.

\begin{figure*}[ht]
\resizebox{\hsize}{!}{\includegraphics{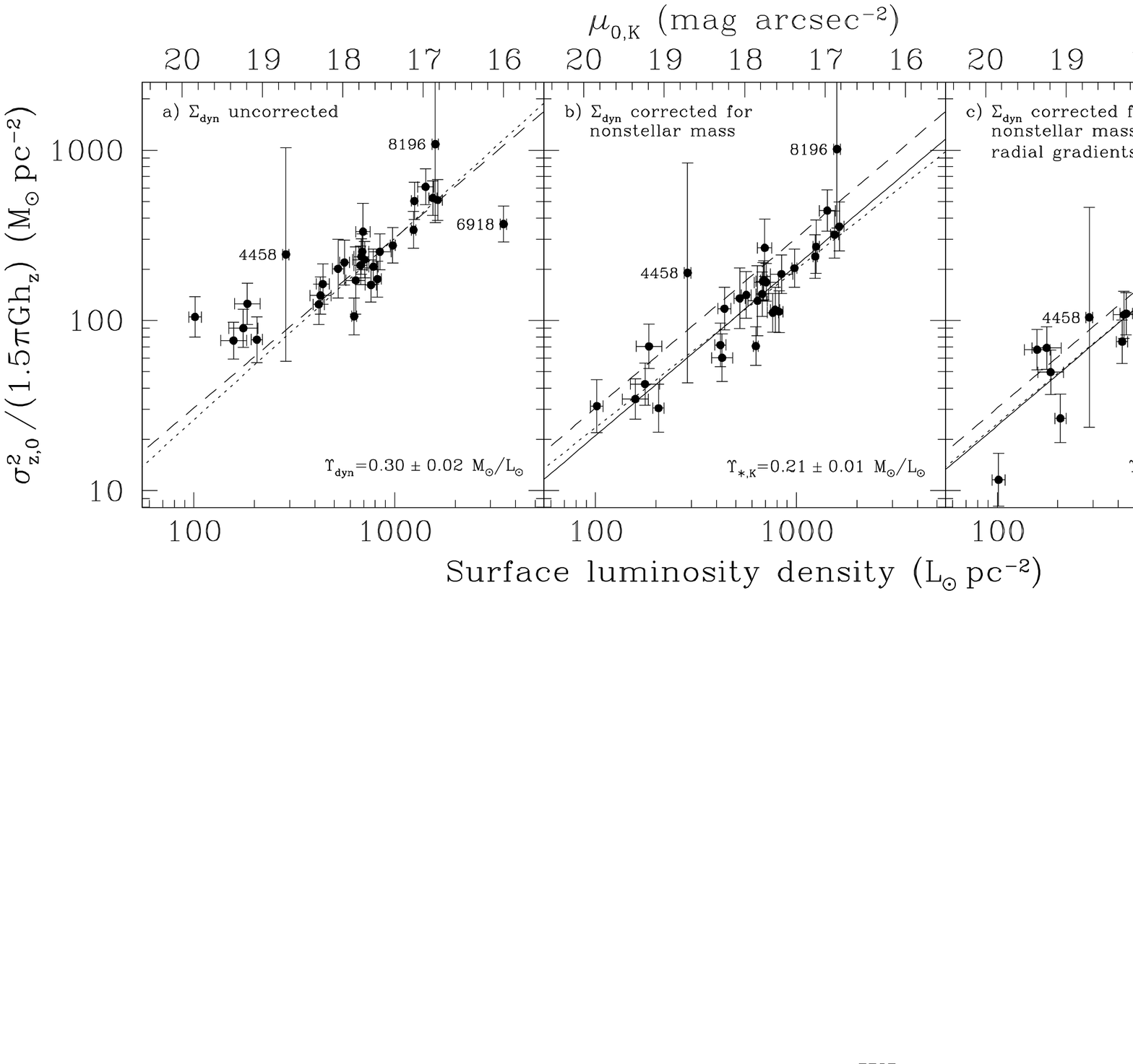}}
\caption{ The \musigrel\ is the relation of surface mass density,
  derived with equation~\ref{eqsigmaz}, with surface luminosity
  density, converted from the $K$-band surface brightnesses assuming
  $M_{\odot,K}=3.30$. Panel a) shows the \musigrel\ for the dynamical
  disk surface mass density \sdyn. In panel b), \sdyn\ is corrected
  for the contribution of gas and dark matter to the disk, and in
  panel c), \sdyn is also corrected for the effects of radial
  gradients in \mlstarK. The long-dashed line in panel a) and solid
  lines are fits with a slope fixed to -0.4, the dotted lines have a
  free slope. The long-dashed lines in panels b) and c) give the
  \musigrel\ from panel a) for reference.}
\label{figmusig}
\end{figure*}

In this Letter, we present the correlation between the central
extrapolated surface brightness \mun\ and the surface mass density
derived from equation~\ref{eqsigmaz}, based on the central
extrapolated vertical velocity dispersion \szn, suggesting
equation~\ref{eqsigmaz} not only applies locally within individual
disks, but also describes the properties of disks across different
galaxies. 

\penalty-10000

\section{Sample and data reduction}

The complete DiskMass sample and its selection is described in detail
DMS-I. Here, we use the sample from DMS-VI, consisting of 30 galaxies
for which PPAK integral-field spectroscopic observations are
available. The galaxies in this sample span a range in properties (see
DMS-VI): Hubble types from Sa to Im (but 83\% have Hubble types Sbc,
Sc, or Scd), absolute magnitudes in broadband $K_s$ (hereafter $K$)
from $M_K=-25.5$ to $M_K=-21$, $B-K$ colors from 2.7 to 4.2 mag, and
$K$-band central disk surface brightnesses from 16 to 20 mag
arcsec$^{-2}$.

We use the observed line-of-sight velocity dispersions derived as
described in DMS-VI. To derive the vertical component of the velocity
dispersion \sz, we follow the same procedure as presented in DMS-VI.
For this, we need to assume a triaxial shape of the stellar velocity
dispersion ellipsoid (SVE). We assume $\alpha=\sz/\sigma_R=0.6\pm
0.15$ and $\beta=\sigma_\theta/\sigma_R=0.7\pm 0.04$, following
DMS-II. This correction for the SVE shape is done for each measurement
of \slos\ in the fibers individually. An exponential function is then
fit to all the derived \sz\ points, excluding radii affected by the
bulge (see DMS-VII). From this fit, we obtain $\szn$ and \hsig, the
dispersion scale length, and their uncertainties.

We determined \hz\ for the galaxies in our sample from their \hR\ 
(from DMS-VI), using the relation for the oblateness parameter
$q=\hR/\hz$ presented in DMS-II, which we estimated to have an
uncertainty of about 25\%.

\section{Results}

Assuming there are no radial gradients in $\alpha$, $\beta$, $k$, \hz,
and \mldyn, equation~\ref{eqsigmaz} can be rewritten in terms of
\szn\ and \mun:
\begin{equation}
\log \sdynn = \log \left(\szn^2/(1.5 \pi G \hz)\right) = -0.4\mun + \log \mldynK.
\end{equation}
Note that \szn\ and \mun\ are not measured in the center, but have
been derived from radial fits to the entire disks (excluding parts
affected by bulges) and are therefore representative of the entire
disk.  Unless otherwise specified, we assume $k=1.5$ (i.e., an
exponential distribution).

\subsection{The dynamical mass-to-light ratio}

In Figure~\ref{figmusig}a, the extrapolated central surface brightness
\mun\ is plotted against the extrapolated central surface mass density
\sdynn.  Most of the points on this \musigrel\ form a well-defined
correlation, but with an unexpectedly small scatter, given that
$\alpha$, $\beta$, \mldynK, $k$, and $q$ could be different from
galaxy to galaxy. There are, however, some outliers. UGC~4458 and
UGC~8196 are early-type galaxies with large bulges, leaving only a few
points in the \sz\ profile that appear unaffected by the bulge, making
our derived \szn\ uncertain. UGC~6918 is a very high surface
brightness (HSB) galaxy that is much more luminous than expected from
the Tully-Fisher relation. Finally, all galaxies at the low surface
brightness (LSB) end, labeled in Figure~\ref{figmusig}a with open
circles, fall above the correlation outlined by the brighter galaxies.

Focusing on the remaining 22 galaxies for the moment, we find a slope
in the \musigrel\ of $-0.43\pm 0.05$ (dotted line in
Figure~\ref{figmusig}a). This is statistically indistinguishable from
the $-0.4$ slope expected for a linear correlation between the surface
mass density and the surface luminosity density. We therefore adopt a
slope of $-0.4$ and fit again (dashed line).  Assuming
$M_{\odot,K}=3.30$ (following Westfall \etal\ 2011; hereafter DMS-IV),
we find that the average \mldynK\ is $0.30\pm 0.02$ \mlunit. The
scatter about the \musigrel\ is $0.11\pm 0.02$~dex.

In DMS-VII, we found $\langle\mldynK\rangle=0.39$~\mlunit. The
difference between that result and the one presented above is mainly
due to the 8 galaxies excluded here. Including all galaxies, we find
$\langle\mldynK\rangle=0.40$~\mlunit, in excellent agreement with our
earlier result.

\subsection{The stellar mass-to-light ratio}

To derive \mlstarK\ from \mldynK, we need to correct for the
contribution of nonstellar mass in the disk. Both \mun\ and \szn\ were
derived from a fit over a large range of the exponential disk.  To
determine the contribution of the gas, we used a similar method,
fitting an exponential to the gas distribution between 0.5 and 3 disk
scale lengths and extrapolating that to the center. For UGC~6918,
already excluded, this correction is larger than \sdyn, which we
attribute to uncertainties in the derived molecular gas mass (see
DMS-VII).  After correcting for the contribution of the gas, we find
$\langle\mlstarK\rangle=0.26\pm 0.02$ \mlunit, and the scatter is
$0.12\pm 0.02$~dex.

As mentioned above, the galaxies at the LSB end tend to fall above the
\musigrel, even after correction for the contribution of the gas. LSB
galaxies are dominated by dark matter for plausible values of
\mlstar\ (e.g., Swaters \etal\ 2003; Kuzio de Naray \etal\ 2008). The
effect of the dark halo on the stellar dynamics could therefore be
non-negligible, as we calculated at the HSB end for UGC~463 (DMS-IV),
and larger at the LSB end. To estimate the effect of the dark matter
on our results, we used previous work by Bottema (1993). In his
Figure~15, Bottema shows the correction to \szn\ that is due to the
dark matter, as a function of $\varepsilon$, the ratio of dark to
stellar density in the midplane. We calculate
$\rho_\mathrm{DM}(r,z=0)$ from the best fit pseudo-isothermal halo
model from DMS-VII. The mid-plane density for the disk is calculated
assuming an exponential vertical density distribution, and
\mldynK\ derived above. Within each galaxy, the value of $\varepsilon$
is roughly constant between one and three disk scale lengths (see also
Figure~17 in DMS-IV); we use the ratio at two disk scale lengths
because it is representative of the radial range over which we fit
\sz\ and $\mu_K$. Among galaxies, $\varepsilon$ ranges from 0.15 for
galaxies at the HSB end to about 1 at the LSB end. With this ratio,
and Bottema's curve, we corrected the values of \szn\ for the effect
of the dark halo.

After \szn\ is corrected for both the contribution of gas and dark
matter, the LSB galaxies follow the same correlation (see
Figure~\ref{figmusig}b). Left free in the fit, the slope is $-0.37\pm
0.03$ (dotted line). Fixing the slope to $-0.4$ (solid line), we find
that $\langle\mlstarK\rangle$, now including the LSB end, is $0.21\pm
0.01$~\mlunit.

The scatter about the correlation has increased to $0.13\pm 0.02$ dex,
due to the uncertainties associated with the corrections for gas and
especially the dark matter. For example, the mass models in DMS-VII
used individual \mlstarK\ for each galaxy, whereas here we use an
average value, but this effect is small because the dark matter
dominates. In addition, we should have iterated the mass modeling
because changing \mlstarK\ will change the halo parameters, which in
turn change \mlstarK. Tests on individual galaxies indicate this may
lower \mlstarK\ another 30\%. We will revisit the issue of the
influence of the dark halo on the disk in a forthcoming paper.

Even though the correction for the contribution of dark matter is
uncertain, it is clear that the correction is larger for galaxies with
lower surface brightness. With the above method, the correction at the
HSB end is about 10\%, and at the LSB end it is about 50\%.

\subsection{The impact of radial gradients}

Above, we assumed that there are no radial variations in $\alpha$,
$\beta$, $k$, $q$, and \mldyn\ within each galaxy. If there are no
gradients, then from equation~\ref{eqsigmaz} it follows that
$\hsig=2\hR$.  In DMS-VI, we found that the ratio
$\log(\hsig/2\hR)=0.07\pm 0.09$, indicating that there is no
significant deviation from this expectation on average. However, there
is some scatter, suggesting that in some galaxies radial gradients may
be present.

Within galaxies, variations in $\beta$ with radius are not expected to
have much impact due to the near-face-on nature of our sample and the
small expected range in $\beta$ (e.g., DMS-II). Within galaxies, there
is little or no radial variation in \hz, at least for late-type
galaxies (e.g., de Grijs \& Peletier 1997; Bizyaev \& Mitronova
2002). Simulations suggest that $\alpha$ is also relatively constant
with radius within the disks of late-type disk galaxies (e.g., Minchev
\etal\ 2012). Radial variations within galaxies are therefore likely
dominated by variations in $\mlstarK$ and $k$. Variations in $k$ could
be caused by changes in the relative contributions of stars, gas and
dark matter.

If we assume that radial gradients are dominated by changes in
\mldyn\ and that both the surface brightness profile and the
\sz-profile have an exponential decline, then the effect of a radial
gradient can easily be estimated. In that case, from
equation~\ref{eqsigmaz}, we find that
\begin{equation}
\mldyn(R)=\mldynn\, e^{(r/\hR)(2-H)/H},
\label{equpsilon}
\end{equation}
where $H=\hsig/\hR$. For $H=2$, $\mldyn(R)$ is constant with radius,
as expected. For other values of $H$ this is not the case, but there
will be some radius $r_a$ for which $\mldyn(r_a)$ is representative of
the average $\mldyn$ across the measured range. This radius $r_a$ is
different from galaxy to galaxy, but on average $r_a=1.0h$. With
equation~\ref{equpsilon} for $r=1.0h$, we find that the correction
factor is $e^{(2-H)/H}$.

Applying this correction converts $\mldynn$ to
$\langle\mldyn(R)\rangle$, which reduces the scatter in the \musigrel,
as shown in Figure~\ref{figmusig}c.  Left free in the fit, the slope
is $-0.39\pm 0.03$ (dotted line). Fixing the slope to -0.4 (solid
line), we find $\langle\mlstarK\rangle = 0.24\pm 0.01$~\mlunit. The
overall scatter remains $0.13\pm 0.02$ dex.  At the LSB end the
scatter is higher (0.2 dex), likely because \hsig\ cannot be measured
as accurately at the LSB end. In addition, for the LSB galaxies the
contribution of dark matter at large radii increases, which can change
the effective $k$. However, this effect is already corrected for in
the dark matter correction above, meaning that the LSB galaxies may be
overcorrected. Considering the same 22 galaxies as above, the scatter
is reduced to $0.09\pm 0.02$ dex, smaller than for the uncorrected
\musigrel.

\subsection{Intrinsic scatter}

There are three main sources of scatter on the \musigrel.  One source
is the uncertainties on the adopted parameters $\alpha$, $\beta$, and
$q$. Our adopted scatter of 0.15 in $\alpha$ between galaxies
introduces a scatter of 0.11 dex on the \musigrel\, and the 25\%
uncertainty on $q$ from galaxy to galaxy also introduces a scatter of
0.11 dex. However, variations in $q$ may be coupled to variations in
$\alpha$, because, at a given \mldyn, galaxies with larger \hz\ will
have higher \sz\ as well. Such a coupling, the details of which depend
on the in-plane heating of \sR, would lessen the impact of
variations in $\alpha$ and $q$ on the scatter on the \musigrel. Due to
the orientation of the galaxy disks, the impact of uncertainties in
$\beta$ and inclination are small.

The second source is the measurement uncertainties on \slos, \mun, and
\hR. These uncertainties contribute 0.05 dex to the scatter.

The remaining source is the intrinsic scatter in the \musigrel\ (\si),
mainly due to variations in $\mlstar$ and possibly in $k$. To estimate
\si, we compared the measured scatter in the \musigrel\ to the median
uncertainty on \sdyn. For HSB galaxies, after correction for the
contribution of gas, dark matter, and radial gradients, the median
uncertainty is 0.12~dex. This is similar to but somewhat higher than
the measured scatter of 0.09~dex (about 25\%), which could be due to
the correlation between $\alpha$ and $q$ mentioned above. Assuming
that the measured scatter is dominated by uncertainties in $\alpha$
and $q$, \si\ must be small, at most about half the measured scatter,
i.e., 12\%, because otherwise the measured scatter about the
\musigrel\ would have been larger. If we assume instead that $\alpha$
and $q$ do not contribute to the uncertainty on \sdyn, \si\ is 20\%.

\section{Discussion and conclusions}

Our main result is that \mun, the extrapolated central disk surface
brightness, and \sdynn, the extrapolated central disk surface mass
density are tightly correlated for the galaxies in our sample.  With
the \musigrel, the dynamical surface mass density can be predicted
from the surface brightness with an accuracy of about 30\% for
galaxies brighter than $\mun=18.5$ mag arcsec$^{-2}$, but galaxies at
the LSB end fall above this \musigrel. After correcting \sdynn\ for
the contribution of gas, dark matter, and the effects of radial
gradients, the galaxies at the LSB end move onto the \musigrel\ as
well. At the LSB end, the scatter about the \musigrel\ remains larger
than at the HSB end, but at the HSB end the scatter is reduced to
about 25\%.

The small scatter around the \musigrel\ is unexpected, given that many
of the galaxies' properties may contribute to it, in particular the
parameters $\alpha$, $k$, $q$, and \mlstarK. Different galaxies may
have different star formation histories, which are expected to
modulate $\mlstarK$.  Variations in vertical profile shapes (e.g., due
to superthin disks, see Schechtman-Rook \& Bershady 2013) may lead to
variations in $k$ among galaxies.  Dominant disk-heating processes may
be different between galaxies, leading to different $\alpha$.  Any of
these variations would have increased the scatter in the \musigrel.
Given the small scatter in the \musigrel, the variations in these
properties among galaxies must be small.  Larger variations in these
parameters are possible, but only if there is fine-tuning among the
parameters (specifically, $\alpha^2 q/(k\mlstarK)$ should be constant)
to maintain the small scatter in the \musigrel.

If we assume conservatively that $\alpha$ and $q$ correlate, as
described above, \si\ is about 20\% and is dominated by variations
between galaxies in $k$ and \mlstarK. If $\alpha$ and $q$ do not
correlate, variations in $k$ and \mlstarK\ may be as low as 12\%.

Assuming $k=1.5$ for all galaxies, a slope of -0.4 in the \musigrel,
and adopting an intrinsic scatter of 12\%, we find that the average
$\mldynK = 0.30 \pm 0.02$~\mlunit, with an intrinsic scatter of
0.04~\mlunit. If $k$ varies between galaxies, the range in $\mldynK$
may be larger, as long as the product of $k$ and $\mldynK$ remains
constant (see equation.~\ref{eqsigmaz}). After correction for gas and
dark matter, as well as gradients within each galaxy's disk, we find
the average $\mlstarK=0.24\pm0.01$~\mlunit, with an intrinsic scatter
of 0.03~\mlunit. This result suggests that, despite spanning a wide
range in properties, galaxies in our sample have similar \mlstarK,
with only small variations from galaxy to galaxy, as was also found in
DMS-VII.

We compare our dynamically inferred $\Upsilon_*^K$ to two canonical
stellar population synthesis models with known differences in how they
treat late-phases of stellar evolution (Bruzual \& Charlot 2003, BC03;
Maraston 2005, M05). All models predict a range of $\Upsilon_*^K$
depending on age, star-formation and chemical enrichment history.  For
the restricted subset of models with solar metallicities and
exponentially declining star-formation rates with e-folding times
between 0.1 Gyr and $\infty$ and ages above 7 Gyr, both models yield
similar ranges of $0.4 < \Upsilon_*^K < 0.65$, for the mean color of
$g-i=0.88$ of our sample. For younger ages, mimicking galaxies with
more vigorous recent star-formation, $\Upsilon_*^K$ drops to 0.25
(0.33) at 3-7 Gyr and 0.15 (0.26) at 0.8-3 Gyr for M05 (BC03)
respectively.

Our mean \mlstarK\ is compatible with the lower end of the
\mlstarK\ values predicted by BC03 and M05 for rather young ages
(suggesting significant recent star-formation).  Alternatively, our
derived \mlstarK\ would change systematically for different adopted
values for $\alpha$, $q$, or $k$, while the scatter in the \musigrel
would remain the same.  To realize $\mlstarK\sim 0.4$, for example,
changes of around 20\% are needed in each of $\alpha$, $q$, and
$k$. Different adopted initial stellar mass functions would also
modulate \mlstarK\ (e.g., Conroy \etal\ 2009).

We note that our sample is biased towards late-type spiral
galaxies. The two early-type galaxies in our sample fall above the
\musigrel, but are consistent with it within their large uncertainties
on \szn. To verify whether this could play a role, we investigated the
results by Herrmann \& Ciardullo (2009) and Gerssen \& Shapiro Griffin
(2012).  We cannot make direct comparisons because the analyses were
done differently, but we do find that the Sc galaxy in their sample
falls on our \musigrel, and the earlier types fall significantly
above, consistent with what we find here.  This could in part be due
to different $q$; in DMS-II we found $q$ may be about 50\% lower in
early-type disk galaxies. However, this can at best explain a small
fraction of the difference. To explain the offset, \mlstarK\ would
have to increase by a factor of 2 or 3 towards early-type
galaxies. This suggests that the \musigrel\ presented here could be a
slice through a plane in which Hubble type, or a physical property
strongly correlated with Hubble type, is a second parameter.

For the sample presented here, the scatter in the \musigrel\ is small,
with an observed scatter at the HSB end of about 25\%, and an
intrinsic scatter of at most 10\% to 20\%. This means that it is
possible to determine the stellar surface mass density from the
observed surface brightness with an accuracy of 10\% to 20\%. It also
means that $\alpha$, $q$, and $k$ cannot vary significantly within our
sample. Finally, unless $k$ changes from galaxy to galaxy, the small
scatter also means that $\mlstarK$ does not vary more than 10\% to
20\% between the galaxies in our sample, and that the average
\mlstarK\ of the galaxies in our sample is $0.24$~\mlunit,
with an estimated intrinsic scatter of at most 0.05~\mlunit.

\acknowledgments

MAB acknowledges support from NSF/AST-1009471. TPKM acknowledges
support from The Netherlands Research School for Astronomy (NOVA). KBW
acknowledges grants OISE-754437 (NSF) and 614.000.807 (NWO).


\begin{references}

\reference{1984ApJ...276..156B} Bahcall, J.~N.\ 1984, \apj, 276, 156

\reference{2010ApJ...716..198B} Bershady, M.~A., Verheijen, M.~A.~W.,
Swaters, R.~A., et al.\ 2010a, \apj, 716, 198 (DMS-I)

\reference{2010ApJ...716..234B} Bershady, M.~A., Verheijen, M.~A.~W.,
Westfall, K.~B., et al.\ 2010b, \apj, 716, 234 (DMS-II)

\reference{2011ApJ...739L..47B} Bershady, M.~A., Martinsson,
T.~P.~K., Verheijen, M.~A.~W., et al.\ 2011, \apjl, 739, L47

\reference{2002A&A...389..795B} Bizyaev, D., \& Mitronova, S.\ 2002,
\aap, 389, 795

\reference{1993A&A...275...16B} Bottema, R.\ 1993, \aap, 275, 16 

\reference{2013ApJ...779..115B} Bovy, J., \& Rix, H.-W.\ 2013, \apj,
  779, 115

\reference{2003MNRAS.344.1000B} Bruzual, G., \& Charlot, S.\ 2003,
\mnras, 344, 1000 (BC03)

\reference{2009ApJ...699..486C} Conroy, C., Gunn, J.~E., \& White,
M.\ 2009, \apj, 699, 486

\reference{1997MNRAS.290..533D} de Blok, W.~J.~G., \& McGaugh,
  S.~S.\ 1997, \mnras, 290, 533

\reference{1997A&A...320L..21D} de Grijs, R., \& Peletier,
R.~F.\ 1997, \aap, 320, L21

\reference{2011MNRAS.417.1621D} Dutton, A.~A., Brewer, B.~J.,
Marshall, P.~J., et al.\ 2011, \mnras, 417, 1621

\reference{2012MNRAS.423.2726G} Gerssen, J., \& Shapiro Griffin,
K.\ 2012, \mnras, 423, 2726

\reference{2009ApJ...705.1686H} Herrmann, K.~A., \& Ciardullo,
R.\ 2009, \apj, 705, 1686

\reference{2003ApJ...586..143K} Kranz, T., Slyz, A., \& Rix,
H.-W.\ 2003, \apj, 586, 143

\reference{2005MNRAS.358..503K} Kregel, M., van der Kruit, P.~C., \&
Freeman, K.~C.\ 2005, \mnras, 358, 503

\reference{2008ApJ...676..920K} Kuzio de Naray, R., McGaugh, S.~S., \&
de Blok, W.~J.~G.\ 2008, \apj, 676, 920

\reference{2005MNRAS.362..799M} Maraston, C.\ 2005, \mnras, 362, 799 (M05)

\reference{2013A&A...557A.130M} Martinsson, T.~P.~K., Verheijen,
M.~A.~W., Westfall, K.~B., et al.\ 2013a, \aap, 557, A130 (DMS-VI)

\reference{2013A&A...557A.131M} Martinsson, T.~P.~K., Verheijen,
M.~A.~W., Westfall, K.~B., et al.\ 2013b, \aap, 557, A131 (DMS-VII)

\reference{2012A&A...548A.126M} Minchev, I., Famaey, B., Quillen,
A.~C., et al.\ 2012, \aap, 548, A126

\reference{1997ApJ...483..103S} Sackett, P.~D.\ 1997, \apj, 483, 103

\reference{2013ApJ...773...45S} Schechtman-Rook, A., \& Bershady,
M.~A.\ 2013, \apj, 773, 45

\reference{2003ApJ...583..732S} Swaters, R.~A., Madore, B.~F., van den
Bosch, F.~C., \& Balcells, M.\ 2003, \apj, 583, 732

\reference{1985ApJ...295..305V} van Albada, T.~S., Bahcall, J.~N.,
Begeman, K., \& Sancisi, R.\ 1985, \apj, 295, 305

\reference{2001ApJ...546..931W} Weiner, B.~J., Sellwood, J.~A., \&
Williams, T.~B.\ 2001, \apj, 546, 931

\reference{2011ApJ...742...18W} Westfall, K.~B., Bershady, M.~A.,
Verheijen, M.~A.~W., et al.\ 2011, \apj, 742, 18 (DMS-IV)

\end{references}
\end{document}